\documentclass[conference]{IEEEtran}
\IEEEoverridecommandlockouts
\usepackage{amsthm}

\usepackage{microtype}
\usepackage{amsmath}
\usepackage{amsfonts}
\usepackage{amssymb}
\usepackage{tabu}
\usepackage{url}
\usepackage{graphicx}
\usepackage{booktabs}

\usepackage{xspace}
\usepackage{multirow}
\usepackage{enumitem}
\usepackage{amsmath}
\usepackage{mathtools}

\newcommand{\ms}[1]{\ensuremath{\mathsf{#1}}}

\newcommand{\Z}{\mathbb{Z}}

\newcommand{\TT}{\ms{T}}

\usepackage{tikz}

\usepackage{cite}
\usepackage{amsmath,amssymb,amsfonts}
\usepackage{algorithmic}
\usepackage{graphicx}
\usepackage{textcomp}
\usepackage{xcolor}
\usepackage{multirow}
\usepackage{subcaption}
\usepackage{booktabs, siunitx}
\usepackage{balance}
\def\BibTeX{{\rm B\kern-.05em{\sc i\kern-.025em b}\kern-.08em
    T\kern-.1667em\lower.7ex\hbox{E}\kern-.125emX}}
\begin{document}

\title{Presto: Hardware Acceleration of Ciphers for Hybrid Homomorphic Encryption\\
{

}}

\author{\IEEEauthorblockN{Yeonsoo Jeon}
\IEEEauthorblockA{\textit{Electrical and Computer Engineering} \\
\textit{The University of Texas at Austin}\\
Austin, USA \\
yeonsoo@utexas.edu}
\and
\IEEEauthorblockN{Mattan Erez}
\IEEEauthorblockA{\textit{Electrical and Computer Engineering} \\
\textit{The University of Texas at Austin}\\
Austin, USA \\
mattan.erez@utexas.edu
}
\and
\IEEEauthorblockN{Michael Orshansky}
\IEEEauthorblockA{\textit{Electrical and Computer Engineering} \\
\textit{The University of Texas at Austin}\\
Austin, USA \\
orshansky@utexas.edu
}

}

\maketitle

\begin{abstract}
Hybrid Homomorphic Encryption (HHE) combines symmetric key and homomorphic encryption to reduce ciphertext expansion crucial in client-server deployments of HE. Special symmetric ciphers, amenable to efficient HE evaluation, have been developed. Their client-side deployment calls for performant and energy-efficient implementation, and in this paper we develop and evaluate hardware accelerators for the two known CKKS-targeting HHE ciphers, HERA and Rubato. 

We design vectorized and overlapped functional modules. The design exploits transposition-invariance property of the MixColumns and MixRows function and alternates the order of intermediate state to eliminate bubbles in stream key generation, improving latency and throughput. We decouple the RNG and key computation phases to hide the latency of RNG and to reduce the critical path in FIFOs, achieving higher operating frequency.

We implement the accelerator on an AMD Virtex UltraScale+ FPGA. Both Rubato and HERA achieve a 6× improvement in throughput compared to the software implementation. In terms of latency, Rubato achieves a 5× reduction, while HERA achieves a 3× reduction. Additionally, our hardware implementations reduce energy consumption by 75× for Rubato and 47× for HERA compared to their software implementation.
\end{abstract}


\section{Introduction}
\label{Introduction}
Homomorphic Encryption (HE) is an encryption scheme that allows computation directly on ciphertexts. This capability enables privacy-preserving computation in client-server settings, making it attractive for many applications, particularly in health and finance. Typically, the client encrypts sensitive data into ciphertext and transmits it to the server for processing. However, despite its privacy advantages, existing HE schemes face two primary challenges. First, the ciphertext-to-plaintext ratio is high, resulting in significant bandwidth requirements for ciphertext transmission. Second, performing HE encryption on resource-constrained client devices incurs substantial computational costs.

Recently, Hybrid Homomorphic Encryption (HHE) has been introduced by \cite{hera, rubato, pasta, sasta} to overcome these challenges. HHE integrates symmetric cryptography with HE, whereby the client encrypts messages using symmetric key encryption, and the server homomorphically decrypts and transciphers  (converts) the ciphertext into the HE domain. 

This approach offers two key advantages. First, relying on a symmetric cipher to encrypt the message results in smaller ciphertext, reducing the transmission bandwidth requirement. Second, symmetric encryption is less computationally taxing, making it more suitable for client-side implementation. Consequently, HHE achieves higher performance and lower energy consumption compared to traditional HE \cite{hera, rubato}. Among the proposed HHE frameworks, HERA and Rubato \cite{hera, rubato} are notable for their support of CKKS \cite{ckks}, an HE scheme specifically designed for approximate computations over ciphertexts. The CKKS scheme's ability to handle real-valued numbers makes it especially suitable for privacy-preserving machine learning (ML) applications.
Client-side energy efficiency is critical for the practical adoption of HHE, particularly for compute- and battery-constrained edge processors. 

\emph{In this paper, we evaluate the potential for hardware acceleration of CKKS-compatible schemes by developing the first hardware accelerators for HERA and Rubato.}
We design vectorized and overlapped functional modules, allowing multiple computations to occur in parallel across pipeline stages, thus reducing execution bottlenecks. We leverage the transposition-invariance property of the MixColumns and MixRows operations to allow the system to alternate between row-major and column-major data orderings, eliminating pipeline stalls. We implement RNG decoupling, a technique that separates the sampling of random constants from the main computation, allowing both to proceed in parallel. This not only hides the latency of random number generation but also reduces the size and criticality of FIFOs, permitting higher clock frequencies. We replace expensive multipliers in the matrix-vector computation with shift-and-add logic, taking advantage of the constant, small-coefficient matrix used in these operations to minimize area and critical path delays.

We implement the accelerators for HERA and Rubato using the AMD Virtex UltraScale+ VCU118 FPGA. Compared to their respective software implementations, our hardware implementations improve throughput by 6× (both Rubato and HERA), reduce latency by 5× (Rubato) and 3× (HERA), and decrease energy consumption per key generation by 75× (Rubato) and 47× (HERA). Rubato achieves higher throughput and lower latency than HERA with only modestly higher area, even though in software HERA outperforms Rubato in latency.

In summary, this paper makes the following contributions:
\begin{itemize}
\item We developed the first hardware accelerators for CKKS-compatible HHE schemes, HERA and Rubato.

\item We exploited the transposition-invariance property of the MixColumns and MixRows functions, eliminating the bubbles in the stream key generation, enhancing throughput and reducing latency compared to the naively parallelized design.

\item We decoupled the random number generation and stream key generation phases, effectively masking RNG latency and improving throughput.

\item We demonstrated substantial performance gains over an optimized AVX2 software implementation.

\item We found that, in hardware, Rubato is more performant and energy-efficient than HERA.
\end{itemize}

The rest of the paper is organized as follows: Section \ref{The RtF Transciphering Framework} introduces the RtF transciphering framework used in CKKS-compatible HHE, Section \ref{CKKS-Friendly Symmetric Ciphers} describes the HHE schemes, Section \ref{Design} presents our proposed architecture, Section \ref{Results} discusses the implementation results, and Section \ref{Conclusion} concludes the paper.

\section{Real-to-Finite Transciphering for HHE}
\label{The RtF Transciphering Framework}
Hybrid homomorphic encryption in the context of CKKS has been studied through the perspective of the Real-to-Finite (RtF) transciphering framework \cite{hera}. RtF integrates CKKS \cite{ckks} and FV \cite{fv} schemes to enable efficient homomorphic operations on encrypted real-valued data. In this framework, the client encrypts its privacy-sensitive data using an (HE-friendly) symmetric key encryption (SKE) scheme and sends it to the server. The server homomophically decrypts the ciphertext and outputs a CKKS ciphertext. 

RtF hinges on two key insights. First, FV is essential due to its native support of finite-field operations to homomorphically decrypt the ciphertext produced by the client-side symmetric cipher. Second, CKKS and FV share the ciphertext space. Upon receipt of the client-generated symmetric ciphertext, the server scales it up into the FV ciphertext space to homomorphically evaluate the decryption circuit of SKE using FV. Once decryption is complete, RtF invokes CKKS’s $\text{HalfBoot}$  routine to convert the resulting ciphertext into a CKKS ciphertext. ($\text{HalfBoot}$ is a CKKS bootstrapping procedure that is less computationally taxing than full CKKS bootstrapping.)


\section{CKKS-Friendly Symmetric Ciphers}
\label{CKKS-Friendly Symmetric Ciphers}

The RtF framework, in principle, permits the server to homomorphically evaluate any SKE decryption. However, in practice, efficiency is dictated by the SKE's multiplicative depth because HE permits only a fixed number of multiplications before noise becomes intolerable. 
Dealing with a decryption circuit with a higher depth requires an HE instance with higher parameters for the noise budget, which increases the computational cost. 
These considerations have informed the formulation of HERA and Rubato as HE-friendly ciphers with low multiplicative depth in their decryption circuit. In this section, we describe these schemes.

\label{HHE Scheme}
\subsection{HERA}








The first symmetric-key encryption (SKE) scheme designed for RtF is HERA, which employs randomized key scheduling as its design principle. HERA encrypts a real-valued message vector \(\mathbf{m} \in \mathbb{R}^{16}\). Given an input key \(\mathbf{k} \in \mathbb{Z}_q^{16}\), HERA's stream key generation function is defined as:
\[
\text{HERA}(\mathbf{k}) = \text{Fin} \circ \text{RF}_{r-1} \circ \cdots \circ \text{RF}_1 \circ \text{ARK}(\mathbf{k})
\]

\noindent where \(\text{Fin}\) is the final round function, and \(\text{RF}\) is the intermediate round function. The final round function is explicitly defined as:

\[
\begin{aligned}
\text{Fin} &= \text{ARK} \circ \text{MixRows} \circ \text{MixColumns} \\
           &\circ \text{Cube} \circ \text{MixRows} \circ \text{MixColumns}
\end{aligned}
\]

\noindent The intermediate round function, executed iteratively for a total of \( r \) rounds, is given by:

\[
\text{RF} = \text{ARK} \circ \text{Cube} \circ \text{MixRows} \circ \text{MixColumns}.
\]

\noindent Each of these round functions is comprised simpler component operations. A crucial component is the randomized key-scheduling add round key function, \(\text{ARK}\), defined as:

\[
\text{ARK}(\mathbf{x}, \mathbf{k}, \mathbf{rc}) = \mathbf{x} + \mathbf{k} \odot \mathbf{rc}
\]

\noindent where \(\mathbf{x} \in \mathbb{Z}_q^{16}\) is an intermediate state, and \(\mathbf{rc} \in \mathbb{Z}_q^{16}\) is a random constant generated by an extendable output function (XOF) using a given nonce \(\text{nc}\).

\(\text{MixColumns}\) and \(\text{MixRows}\) are linear functions. Each implements a matrix multiplication when the intermediate state $\mathbf{x}$ is treated as a $4 \times 4$ matrix in $\Z_q$. Specifically, if $\mathbf{x} = (x_1, x_2, \cdots, x_{16}) \in \Z_q^{16}$, then we can map it into the matrix $X$: 
\begin{equation}
\mathbf{x} \mapsto
X =
\begin{bmatrix}
x_1  & x_2  & x_3  & x_4  \\
x_5  & x_6  & x_7  & x_8  \\
x_9  & x_{10} & x_{11} & x_{12}\\
x_{13}& x_{14}& x_{15}& x_{16}
\end{bmatrix}
\label{eq:matrix_mapping}
\end{equation}

\noindent Let 

\[
M_v = 
\begin{bmatrix}
2 & 3 & 1 & 1 \\
1 & 2 & 3 & 1 \\
1 & 1 & 2 & 3 \\
3 & 1 & 1 & 2 
\end{bmatrix}
\]

\noindent \(\text{MixColumns}\) multiplies $M_v$ by the columns of $X$, and \ms{MixRows} multiplies $M_v$ by the rows of $X$. For $c\in \{0, 1, 2, 3\}$, 

\[
\begin{bmatrix}
y_{0,c}\\y_{1,c}\\y_{2,c}\\y_{3,c}
\end{bmatrix}
= M_v
\begin{bmatrix}
x_{0,c}\\ x_{1,c}\\x_{2,c}\\x_{3,c}
\end{bmatrix}, 
\qquad
\begin{bmatrix}
y_{c,0}\\y_{c,1}\\y_{c,2}\\y_{c,3}
\end{bmatrix}
=
M_v 
\begin{bmatrix}
x_{c,0}\\ x_{c,1}\\x_{c,2}\\x_{c,3}
\end{bmatrix}
\]

Finally, the nonlinear function \(\text{Cube}\) is defined as:

\[ \text{Cube}(\mathbf{x}) = (x_1^3, x_2^3, \cdots x_{16}^3)\]
\noindent for $\mathbf{x} = (x_1, x_2, \cdots, x_{16}) \in \Z_q^{16}$. 



\subsection{Rubato}

Rubato retains HERA’s round structure, but introduces several modifications.
First, it replaces a nonlinearity based on $\text{Cube}$ with a Feistel function. In making this choice, Rubato prioritizes efficiency of a lower-degree nonlinear transformation, that allows a much lower multiplicative depth. Second, it modifies the final round function to include addition of discrete Gaussian noise. Third, the length of the input vector is now a cipher design parameter (compared to a fixed vector length of 16 elements in HERA). 
The stream key generation function in Rubato is:
\[  \text{Rubato}(\mathbf{k}) = \text{AGN} \circ \text{Fin} \circ \text{RF}_{r-1} \circ \cdots \circ \text{RF}_{1} \circ \text{ARK}(\mathbf{k})\]

\noindent The final round function is defined as:
\[
\begin{aligned}
\text{Fin} &= \text{Tr} \circ \text{ARK} \circ \text{MixRows} \circ \text{MixColumns}\\
         &\circ \text{Feistel} \circ \text{MixRows} \circ \text{MixColumns}
\end{aligned}
\]

\noindent The round function is defined as:
\[ 
\text{RF} = \text{ARK} \circ \text{Feistel} \circ \text{MixRows} \circ \text{MixColumns}
\]


ARK in Rubato is identical to HERA except that the length $n$ of the input $\mathbf{x}\in\mathbb{Z}_q^{n}$ is specified by the user. MixColumns and MixRows operate in the same way as in HERA, multiplying $M_v$ by the columns and rows of the state matrix, respectively.
with one major difference. In HERA, the intermediate state is always of size $n=16$, with the corresponding matrix  always being a $4\times4$ matrix. As discussed, in Rubato, the state dimension $n\in\{16,36,64\}$ depends on the chosen cryptographic parameters. Therefore, the intermediate state is reshaped into a $v\times v$ matrix with $v=\sqrt{n}$, making the dimensions of the mixing matrix $M_v$ dependent on the user parameters. 

The non-linear transformation for $\mathbf{x} = (x_1, x_2, \cdots, x_{n}) \in \Z_q^{n}$, based on the Feistel function, is defined as:

\[ \text{Feistel}(\mathbf{x}) = (x_1, x_2+x_1^2, \cdots, x_n+x_{n-1}^2)\]

\noindent Truncation, denoted by Tr, discards the last $n-l$ elements of a vector. For $\mathbf{x}=(x_{1},x_{2},\cdots,x_{n})\in\mathbb{Z}_{q}^{n}$ it is defined by

\[ \text{Tr}_{n, l}(\mathbf{x}) = (x_1, \cdots, x_l)\]

\noindent AGN, that adds discrete Gaussian noise to the input vector, is the final layer. Let $\mathbf{x}=(x_{1},x_{2},\cdots,x_{l})\in\Z_q^{l}$ and let $\mathbf{e}=(e_{1},e_{2},\ldots,e_{l})$, where each $e_{i}$ is sampled independently from a discrete Gaussian distribution. AGN is defined as

\[ \text{AGN}(\mathbf{x}) = (x_1+e_1, x_2+e_2, \cdots, x_l+e_l)\]

\section{Design}
\label{Design}

\begin{figure*}[t!] 
  \centering
  
  \begin{subfigure}[t]{\textwidth}
  \centering
  \includegraphics[width=0.8\textwidth]{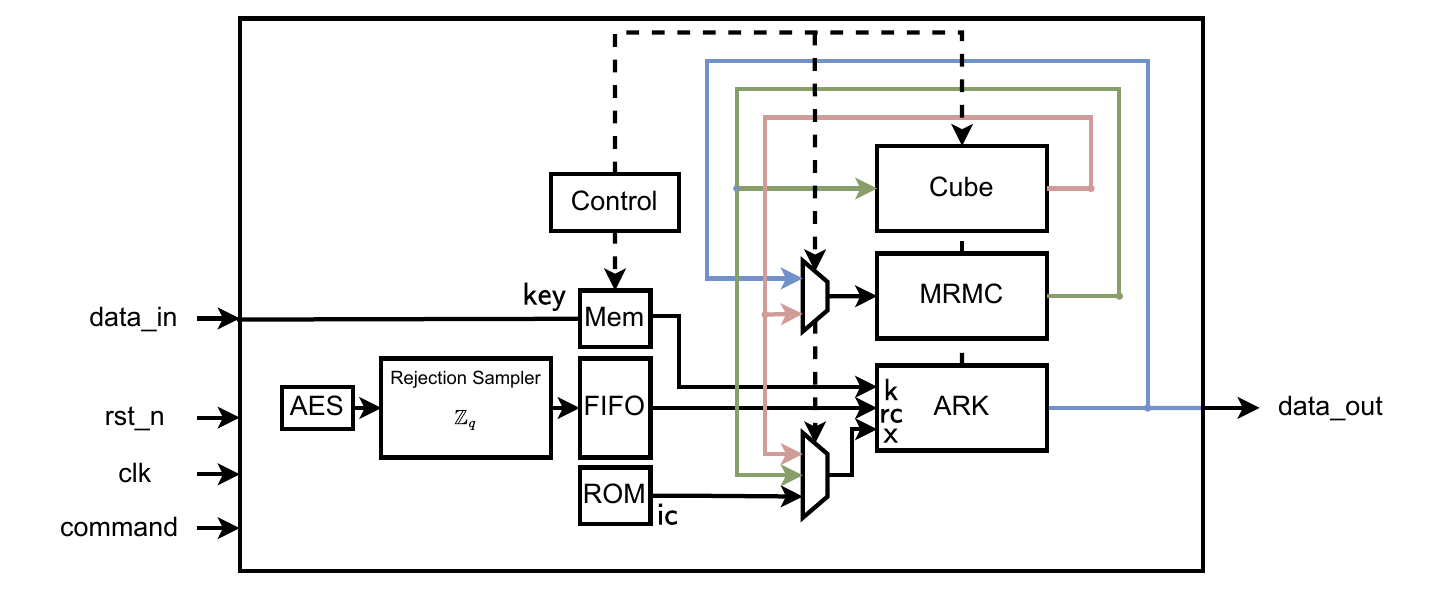}
  \caption{HERA}
  \label{fig:architecture hera}
  \end{subfigure}

  \begin{subfigure}[t]{\textwidth}
  \centering
  \includegraphics[width=\textwidth]{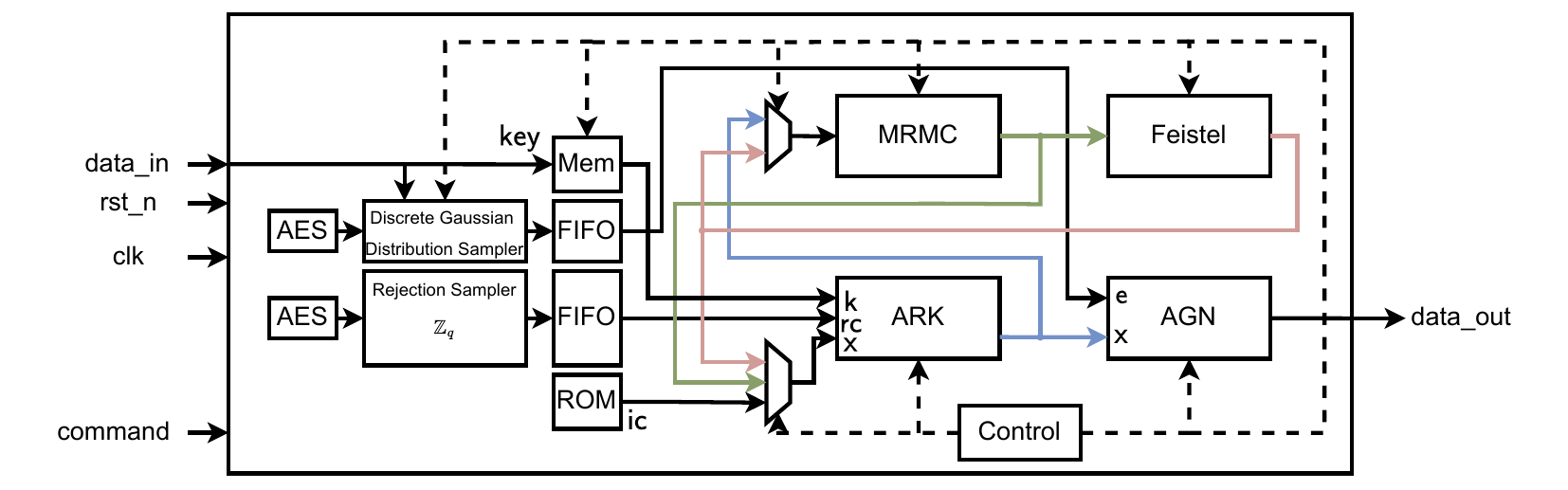}
  \caption{Rubato}
  \label{fig:architecture rubato}
  \end{subfigure}

\caption{Block diagrams of the two accelerators.}
\end{figure*}

\begin{figure*}[t!]                    
  \centering                          

  \begin{subfigure}[t]{0.9\textwidth}
    \includegraphics[width=\textwidth]{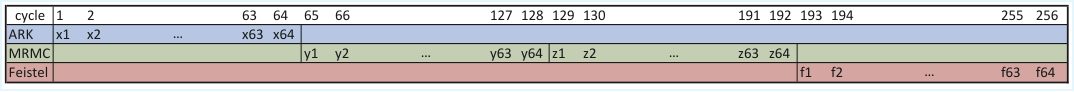}
    \caption{Data schedule for the baseline (non-vectorized) design, showing the output of each module.}
    \label{fig:dataflow_baseline}
  \end{subfigure}
  \hfill                                

  \begin{subfigure}[t]{0.9\textwidth}
  \centering
    \includegraphics[height=0.22\textheight,keepaspectratio]{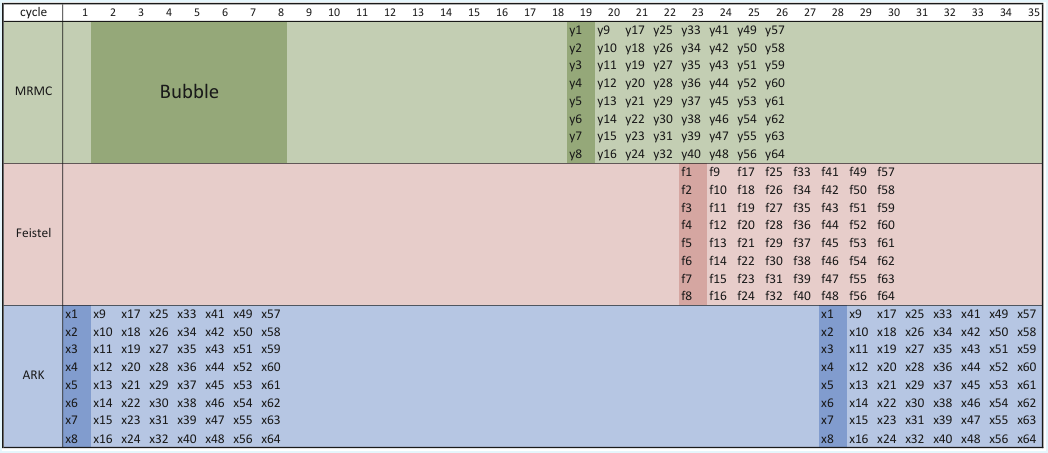}
    \caption{Data schedule for a naively vectorized design. The first row of the intermediate state is highlighted. The state enters RF in row-major order, generated by the previous-round ARK, flowing to MRMC (RF's first layer).
    MRMC stalls until the column of the state is ready (shown as a bubble). 
     MRMC module computes the \ms{MixColumns}, transposition, and \ms{MixRows}, 
     and is followed by Feistel and ARK that all output in row-major order. Same data schedule is repeated in remaing RF layers.}
    \label{fig:dataflow_row_bubble}
  \end{subfigure}
  \hfill

  \begin{subfigure}[t]{0.9\textwidth}
  \centering
    \includegraphics[height=0.22\textheight,keepaspectratio]{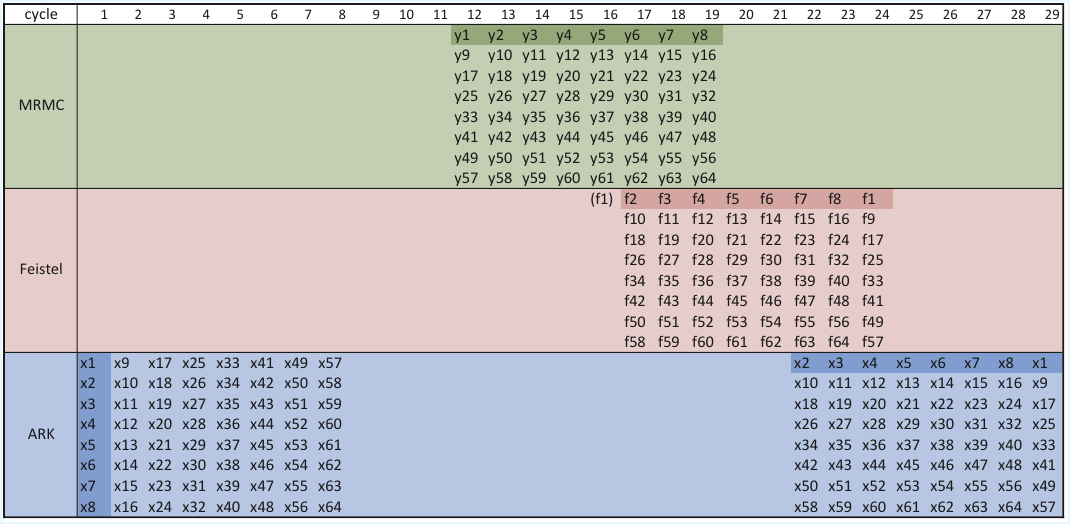}
    \caption{Data schedule after the MRMC optimization that eliminates the bubble. Now the state enters RF in row-major order but MRMC's output is in column-major order. Feistel and ARK retain column-major order, requiring a modified schedule for the next RF layer, see Figure \ref{fig:dataflow_col}. 
    (Note that due to MRMC implementation's state dependency Feistel  stalls.) 
    }
    \label{fig:dataflow_row}
  \end{subfigure}
  \hfill
  
  \begin{subfigure}[t]{0.9\textwidth}
  \centering
    \includegraphics[height=0.22\textheight,keepaspectratio]{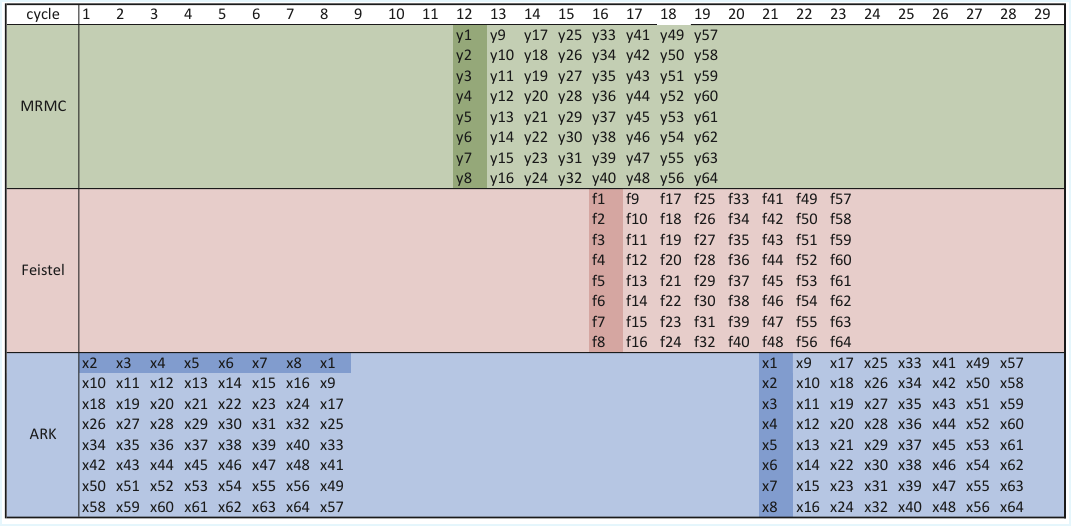}
    \caption{Data schedule when state enters RF in column-major order. MRMC 
    produces a row-major order output, that allows Feistel to proceed without stalling. The state exits RF in row-major order with the next RF layer's schedule alternating back to Figure \ref{fig:dataflow_row}.}
    \label{fig:dataflow_col}
  \end{subfigure}
  \hfill


  \caption{Data schedules of the RF layers in different variants of Rubato implementation. The fully optimized design alternates between row-major and column-major orders of the state. It eliminates the bubble before MRMC, improving latency.
  }
  \label{fig:dataflow}
\end{figure*}

\begin{figure*}[t!]                    
  \centering

  \begin{subfigure}[t]{0.9\textwidth}
  \centering
    \includegraphics[height=0.22\textheight,keepaspectratio]{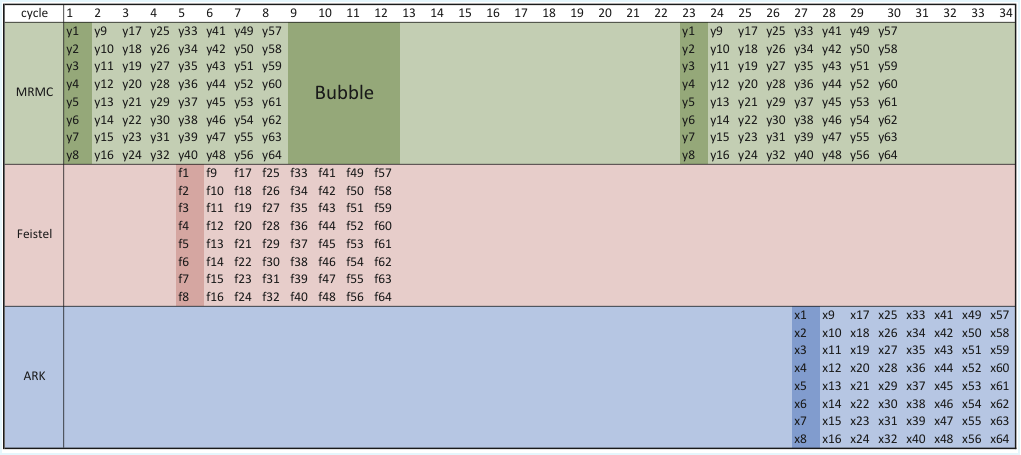}
    \caption{
    Data schedule for a naively vectorized design. Similar to \ref{fig:dataflow_row_bubble}, the second MRMC stalls until the column of the state is available. MRMC produces a row-major output.
}
    \label{fig:dataflow_fin_bubble}
  \end{subfigure}

  \begin{subfigure}[t]{0.9\textwidth}
  \centering
    \includegraphics[height=0.22\textheight,keepaspectratio]{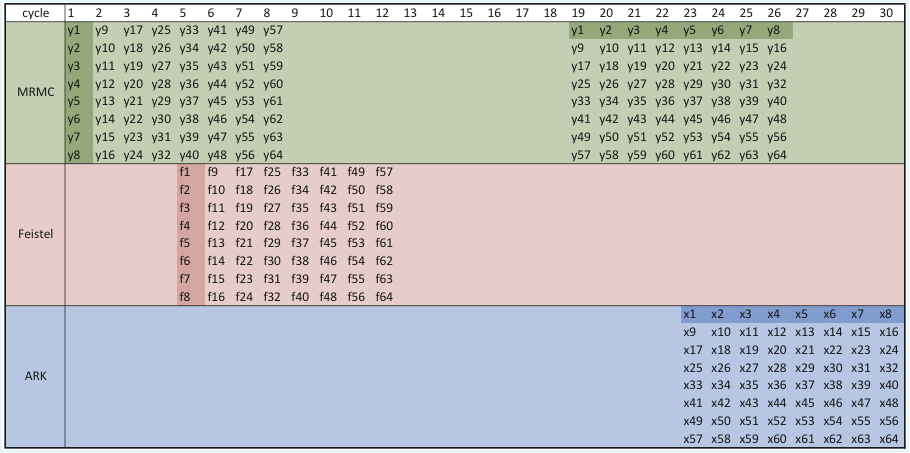}
    \caption{
    Data schedule after the MRMC optimization. The bubble before the second MRMC is eliminated. MRMC now produces a column-major output.
    }
    \label{fig:dataflow_fin}
  \end{subfigure}

  \caption{Data schedules of Fin layers in different variants of Rubato implementation. The fully optimized design eliminates the bubble before the second MRMC, improving latency.
  }
  \label{fig:dataflow}
\end{figure*}

\subsection{Overview}



At high level, the accelerators of both HERA and Rubato instantiate dedicated hardware to realize the transformations needed for cipher computation. 
Figure \ref{fig:architecture hera} and Figure \ref{fig:architecture rubato} show the architectures for both accelerators. The two designs share major building blocks, including ARK, a fused MixColumns/MixRows block MRMC, a Rejection Sampler, and a RNG based on AES. The nonlinear blocks are different: Cube for HERA and Feistel for Rubato. Rubato requires a DGD (Discrete Gaussian Distribution) Sampler (that takes AES output as its random source) and the AGN blocks. A global controller manages the flow.

At initialization, the key is loaded from memory. The Rejection Sampler module samples a random number from $\Z_Q$, required for ARK. The ARK module receives either the constant vector $\mathbf{ic}$, or the output of MixRows, or the output of the nonlinear layer. During the stream key generation, the controller selects one of the three inputs to the ARK module, chosen according to the current round stage. The sampled noise is fed to the AGN module and added to the intermediate state. 

Consider a baseline (scalar) implementation. Figure \ref{fig:dataflow_baseline} shows the dataflow of the first few stages of the intermediate state.
The baseline design executes sequentially with each intermediate state element computed in order. 
To increase throughput and reduce latency, we propose vectorizing computations within every module while overlapping their execution.






\subsection{Optimizing MixColumns and MixRows Matrix-Vector Multiplication}
We note that MixColumns is always followed by MixRows, and that both multiply the state by the same matrix $M_v$. That enables us to use the same hardware, which we refer to as the MRMC module, to implement both functions. A specific optimization we introduce is a data schedule that alternates the order of the intermediate state streaming to MRMC. 

We focus on Rubato, but HERA is similar. We modify the baseline scalar implementation to allow every module (namely, ARK, MRMC, and Feistel) to produce  $v=8$ elements each cycle with basic vectorization. Vectorization enables modules to execute the function simultaneously in each cycle, as soon as the inputs are produced and buffered. We refer to this approach as function overlapping.

We first describe our terminology with respect to the row-major and column-major orders of the intermediate state. Consider the matrix representation of the intermediate state in Equation \ref{eq:matrix_mapping}.
In row-major order, the state flows into or out of the modules one row per cycle (for example, \{$x_1, x_2, x_3, x_4$\}). In column-major order, the state flows one column per cycle (for example, \{$x_1, x_5, x_9, x_{13}$\}).

The progression of the intermediate state, when it enters in row-major order, through stages in the RF layer, and, separately, in the Fin layer, is illustrated in Figures \ref{fig:dataflow_row_bubble} and \ref{fig:dataflow_fin_bubble}. The figures show the output of each module, with the first \textit{row} of the state being highlighted.

Within RF (Figure \ref{fig:dataflow_row_bubble}), ARK  outputs the state in row-major order. However, MRMC cannot begin execution immediately because the required input is not available. To start processing, MRMC needs a column of the intermediate state, which becomes available only in cycle 8, causing MRMC to stall (shown as a bubble in the figure). Once the input is ready, MRMC begins computation (that includes MixColumns, transposition, and MixRows), and outputs the state in row-major order. It is then followed by Feistel and ARK, both of which retain the state in row-major order; as a result, the same data scheduling pattern is repeated in the remaining RF layers.


A bubble also occurs in the Fin layer, which exhibits the same behavior. As shown in Figure \ref{fig:dataflow_fin_bubble}, the output of the MRMC function is in row-major order, as is the output of the Feistel function. Because of this ordering, the second MRMC operation cannot begin immediately and must stall until the Feistel module completes its computation.

Pipeline bubbles introduced by naive vectorization degrade the architecture’s performance. To overcome this, we leverage the MRMC module’s transposition-invariance property in a more efficient data-scheduling strategy. MRMC is the composition of MixColumns followed by MixRows, and can be written

\begin{align*}
  \text{MRMC}(X) &\coloneq (\text{MixRows} \circ \text{MixColumns})(X)    \\
  &= M_v(M_vX)^\TT                                                  \\
  &= M_vX^\TT M_v^\TT.                
\end{align*}

\noindent Applying the same operation to a transposed input yields

\begin{align*}
  \text{MRMC}(X^\TT) &= M_v (X^\TT)^\TT M_v^\TT \\
                   &= M_v X M_v^\TT,
\end{align*}

\noindent while transposing the output gives

\begin{align}
  (\text{MRMC}(X))^\TT &= (M_v X^\TT M_v^\TT)^\TT \notag  \\
                   &= M_v X M_v^\TT. \label{eq:mrmc property}
\end{align}

because $(ABC)^\TT = C^\TT B^\TT A^\TT$. Therefore, regardless of the order of $X$ streaming in to the module, the output is the same up to the transposition. 

We leverage the described transposition-invariance to remove the bubbles in the RF and Fin layers.
The technique, which we call MRMC optimization, is illustrated in Figure \ref{fig:dataflow_row} and \ref{fig:dataflow_fin}. As the state flows into MRMC in row-major order, the module treats the input as a transposed matrix,
allowing execution without stalling. 
Because of that, MRMC outputs the result in column-major order,  matching the transposed result of a column-major input.
Thus, a single pass through MRMC changes the order of the intermediate state (from a row-major input to column-major output).
This results in following Feistel to stall for a cycle due to the intra-data dependnecy of the first column, for example, $f_9 = y_9 + y_8^2$.
Also, this change requires that a different data scheduling be used in the next the RF layer. 
Figure \ref{fig:dataflow_col} shows the data scheduling in RF when the input is in column-major order.

We now see that in our modified implementation, the intermediate state alternates between row-major and column-major order throughout the stream key generation RF layers. Let us consider in Figures \ref{fig:dataflow_row} and \ref{fig:dataflow_col} an example of two consecutive RFs to observe how the state orientation alternates. We assume the state streams in row-major order, after exiting ARK of the previous RF. MRMC transforms it into column-major order, and Feistel retains it. The second RF, therefore, begins with the column-major input, which, however, starts with the second column (as in Figure \ref{fig:dataflow_col}) due to the stall in Feistel. This is because the first column is computed at the last cycle of Feistel. Since MRMC implements matrix multiplication as a sequence of matrix–vector multiplications, it can process the second column first, while computing MixColumns. When the delayed first column arrives, it is handled last, and the state leaves MRMC in the normal row-major order, returning to the original orientation of Figure \ref{fig:dataflow_row}. (The same technique allows the second MRMC operation in the Fin layer to proceed without stalling, with the modified data schedule shown in  \ref{fig:dataflow_fin}).



We further optimize the MRMC module by noting that $M_v$ is a constant matrix with small coefficients. Constant multiplications are realized with shift and add operations instead of full multipliers. This optimization reduces area usage and shortens the critical path.

\subsection{Hiding Latency of Round Constant Sampling by Decoupling}
\label{decoupling}

Sampling random numbers for ARK occupies a large share of total execution time. Both HERA and Rubato demand extensive round constants, forcing many RNG calls. For example, for Rubato's Par-128L parameter set, ARK executes (round+1) times and needs $N$ round constants per execution ($l$ round constants for the final layer). This requirement translates to about 4700 random bits, or, roughly, 37 AES invocations. The original (software) implementation of Rubato reports that randomized key setup latency is dominated by this sampling phase. The software samples all round constants before initiating stream key generation, contributing to high latency.

We observe, however, that the round constant sampling and stream key generation can be executed concurrently. We call this optimization RNG decoupling. The accelerator uses the AES core with 128b/cycle throughput to continuously fill a FIFO with the freshly-sampled round constants. At the same time, ARK consumes the constants on demand, and computes the intermediate state. As long as the throughput of the AES core with the rejection sampler exceeds the consumption rate, the pipeline never stalls, and the latency of random number generation is effectively hidden.

RNG decoupling offers an additional advantage because it permits the use of a small FIFO. By hiding the latency of round constant sampling, the design lowers the FIFO depth needed for these constants and increases clock frequency. In the baseline design, as in the software implementation, all constants must be stored before processing begins, thus, the FIFO must store all constants needed for a complete stream key generation. The FIFO depth reaches 188, for a single lane, and 1504, when 8 lanes run in parallel, resulting in a large FIFO. With RNG decoupling, only a small FIFO is needed to absorb short-term rate mismatches. The shallower FIFO reduces the FIFO pointer fan-out and shortens the critical path, allowing higher clock frequency.

\subsection{Noise Sampling and XOF}
\label{Noise Sampling and XOF}
Rubtao also requires the addition of noise based on Discrete Guassian Distribution. The sampled noise is fed to the AGN module and added to the intermediate state. 

The original (software) implementation of HERA uses SHAKE256 as the XOF. Depending on the parameter set, Rubato supports either AES or SHAKE256. We use AES for both schemes because an AES-based XOF delivers higher throughput at the same clock frequency. (A recent SHAKE256 core reports 14.7 bits/cycle at 100 MHz \cite{shakeHW}. By contrast, the AES core in \cite{aes} delivers 128 bits/cycle at the same clock frequency.) For the Rubato Par-128L parameters, the round constant sampler must supply about 84 bits/cycle. A single AES instance easily meets this requirement, while multiple SHAKE256 instances would be needed, at high area cost. We implement the Discrete Gaussian Sampler using the inverse-CDF method that takes AES output as its random source. 
The method uses the lookup table, in which the CDF values are stored at $(\lambda/2)$-bits of precision\cite{discretegaussian}, required for $\lambda$ bits of security.

\section{Results}
\label{Results}




\subsection{Performance Comparison}

\begin{table*}[t!]                             
\caption{Performance Analysis: Hera}  
\centering
\resizebox{\textwidth}{!}{                      
\begin{tabular}{l r r r r r r}
\toprule
Implementation & Cycles & Time [\si{\micro\second}] & Throughput [\si{Msps}] 
               & Freq.\,[\si{MHz}] & Power [\si{W}] & Energy [\si{\micro\joule}] \\
\midrule
SW (AVX)              & 4575 & 1.52  & 10.5 & 3000 & 65  & 99   \\
D1: Baseline               &  729 & 13.9  &  9.24 &   52.6 & 3.2 & 43   \\
D2: + Decoupling  &  512 &  2.30 & 55.6 &  222   & 4.3 &  9.9 \\
D3: + V/FO/MRMC       &   90 &  0.540 & 65.8 &  167   & 3.8 &  2.1 \\
\bottomrule
\end{tabular}}
\label{tab:performance comparison hera}                          
\end{table*}

\begin{table*}[t!]                             
\caption{Performance Analysis: Rubato}  
\centering
\resizebox{\textwidth}{!}{                      
\begin{tabular}{l r r r r r r}
\toprule
Implementation & Cycles & Time [\si{\micro\second}] & Throughput [\si{Msps}] 
               & Freq.\,[\si{MHz}] & Power [\si{W}] & Energy [\si{\micro\joule}] \\
\midrule
    SW (AVX)              & 5430 &  1.81 &  33.1 & 3000  & 65  & 120   \\
    D1: Baseline             & 1478 & 39.9  &  12.0 &   37.0 & 3.4 & 140   \\  
    D2: + Decoupling &  800 &  4.40 & 109   & 182 & 4.9 &  21   \\
    D3: + V/FO/MRMC       &   66 &  0.376 & 188  & 175 & 4.1 &   1.6 \\
\bottomrule
\end{tabular}}
\label{tab:performance comparison rubato}                          
\end{table*}

We implement the accelerators on the AMD Virtex UltraScale+ VCU118 FPGA. 
We use Par-128a parameter for HERA and Par-128L parameter for Rubato.
For each scheme, we assume that the basic design approach is a non-vectorized implementation, producing one state element per module per cycle, shown in Figure \ref{fig:dataflow_baseline}. 
We describe the implementation of two design options for each scheme. 
Vectorization of a scalar implementation is the first step. 
Because the state is represented by a matrix, with $v=\sqrt{n}$, it is natural to vectorize the design in a way that respects the dimensions of the state matrix. 
This leads us to choose the vector length to be $v$ = 4 for HERA, and $v$ = 8 for Rubato. 

Our goal is to ultimately compare the two accelerators on the set of empirical performance measures, including latency, throughput, and FPGA resource utilization. 
Selecting design parameters that guarantees that the two accelerators end up with the same value of some particular high-level metric is challenging because of the differences in the security parameters.
We aimed to approximately match the final throughput of the baseline designs of each scheme. We translated this goal into a more easily-quantifiable design decision: we match the throughput of processing the state matrix in both schemes. 
To match the single-lane throughput of the vectorized Rubato design (with vector width of 8), that processes 8 elements per cycle, we allow the optimized HERA design have 2 lanes (each being 4 elements wide). Following the same rationale, we use a version of the basic design (utilizing the the basic scalar design), but also able to process 8 elements per cycle by arranging it in 8 identical lanes. We call this ''the baseline design`` in the discussion below. 

In the Tables I and II, we first explore isolating the impact of RNG decoupling. The design referred to as "Baseline + Decoupling" is the baseline, with the \emph{only} difference being that the RNG and rounds are decoupled (i.e., it is a scalar design with no function overlapping, or MRMC optimization).
The final design option includes all proposed modifications, representing a  vectorized design that incorporates the MRMC optimization and function overlapping, added to the RNG decoupling. 



Tables \ref{tab:performance comparison hera} and \ref{tab:performance comparison rubato} show the characteristics of the accelerators with different optimizations and compare them to the reference software implementation. RNG decoupling results in a very significant improvement in throughput, as well as a sizable impact on latency.
Throughput improvements are due to a combination of two mechanisms.
The first mechanism is latency hiding that reduces latency by about 42\% in HERA and about 85\% in Rubato. 
The second mechanism is the increase in achievable clock frequency due to the reduction in FIFO depth. Because the path from the FIFO read pointer to the FIFO data register is on the critical path, the FIFO depth reduction is impactful. We see that clock frequency is increased by 4× and 5× for HERA and Rubato, respectively. Together, the two effects produce an increase in throughput of 6× for Hera and 9× for Rubato.

Next, we analyze the impact of vectorization and function overlapping. In both HERA and Rubato, the fully optimized design (D3) shows a dramatic reduction in latency and a sizable throughput increase compared to design D2.
The improvements in latency and throughput result from the combination of three mechanisms. The first mechanism (labeled V in the Table) is the expected result of vectorization itself, which reduces latency and boosts throughput by 8×. The second mechanism (labeled FO in the Table) is function overlapping; in the case of  Rubato, it reduces latency from 100 to 83 cycles. The third mechanism (labeled MRMC in the Table) is the MRMC optimization, which eliminates bubbles during stream key generation. In Rubato it reduces latency to 66 clock cycles -- an 26\% improvement.
In total, these three mechanisms reduce latency by 4× and 12× for HERA and Rubato, and improve throughput by 1.2× and 1.7× times for HERA and Rubato, respectively.


We compare the performance of accelerators against optimized software implementations of HERA and Rubato on an Intel i7-9700 running at 3 GHz. To ensure a fair comparison, we modified the reference code to use AES, instead of SHAKE256, as the XOF for HERA, as discussed in Section \ref{Noise Sampling and XOF}. We ran the stream key generation function for 1000 executions, discarded the first 250 runs for cache warmup, and then computed the average latency over the remaining 750 runs. The software implementation leverages Intel AVX2 instructions for optimized performance. Even though our hardware designs run at a much lower clock frequency than the CPU, they still deliver about a 6× improvement in throughput for both schemes, and reduce latency by 3× and 5× for HERA and Rubato, respectively. 
We also compare power and energy consumption against the software implementation. The results show that, in the fully optimized design (D3), power consumption is 17× and 16×, for HERA and Rubato, respectively, lower than in software. The reduction in energy consumption compared to software is 47× and 75× for HERA and Rubato, respectively. 

Finally, we compare the results of the HERA and Rubato implementation across software and hardware. In software, HERA achieves lower latency than Rubato. The primary reason HERA exhibits lower latency is its round constant sampling overhead. HERA requires only 96 round constants for generating one stream key, whereas Rubato needs 188.
In hardware, RNG decoupling reduces the total latency by 80\% for Rubato and by 40\% for HERA. Yet HERA still maintains lower latency, (see Table \ref{tab:performance comparison rubato}), because Rubato performs more computation than HERA: while Rubato has fewer rounds (2 compared to 5 in HERA), it computes on a larger vector length parameter (n = 64 versus 16), which increases the total element count to process.
However, in a fully optimized design (D3), Rubato's latency is lower than that of Hera.
This gain is due to the more efficient utilization of the MRMC module by Rubato. 
In HERA, MRMC stalls for 5 cycles while waiting for the inputs of the next RF. In contrast, in Rubato, it stalls for only 2 cycles, leading to better latency/throughput for Rubato.

\subsection{Resource Utilization}

\begin{table}[t!]                             
\caption{Resource Utilization: Hera}  
\centering
\resizebox{\columnwidth}{!}{
\begin{tabular}{l r r r r}
    \toprule
    Implementation          & LUT    & FF     & DSP  & BRAM \\
    \midrule
    D1: Baseline              & 107479 &  25920 &   16 &   86 \\
    D2: + Decoupling   &  37672 &  12401 &   16 &   86 \\
    D3: + V/FO/MRMC         &  48001 &  14846 &   56 &   86 \\
    \bottomrule
  \end{tabular}}
\label{tab:area hera}                          
\end{table}

\begin{table}[t!]                             
\caption{Resource Utilization: Rubato}  
\centering
\resizebox{\columnwidth}{!}{
\begin{tabular}{lrrrr}
    \toprule
    Implementation          & LUT    & FF     & DSP  & BRAM \\
    \midrule
D1: Baseline               & 273503 & 83583 & 32 & 169 \\  
D2: + Decoupling  & 77526  & 38058 & 32 & 169 \\
D3: + V/FO/MRMC               & 64510  & 24577 & 32 & 336.5 \\
    \bottomrule
\end{tabular}}
\label{tab:area rubato}
\end{table}

Tables \ref{tab:area hera} and \ref{tab:area rubato} compare FPGA resource utilization across different designs. Compared to the baseline, the fully optimized designs for both schemes use fewer FFs and LUTs, primarily due to the reduced FIFO depth enabled by RNG decoupling. In HERA, the LUT and FF usage for FIFO decreases by approximately 3×, while Rubato achieves a 6× reduction. Overall, Rubato requires slightly more LUTs and FFs than HERA but uses fewer DSPs. Despite these small increases, the fully optimized Rubato achieves higher throughput and lower latency, outperforming HERA in both performance and area.

\section{Conclusion}
\label{Conclusion}

HERA and Rubato are CKKS-compatible HHE schemes that reduce the communication overhead in transmitting the ciphertext, and achieve higher performance and lower energy compared to the traditional HE. We developed hardware accelerator for HERA and Rubato. We leverage the transposition-invariance property in MRMC function, and improve latency and throughput by vectorization, function overlapping, and bubble elimination. We decouple the RNG and stream key generation phases to hide the latency of sampling random constant. Our design achieve 6× improvement in throughput for both HERA and Rubato and achieve 5×, 3× latency reduction in Rubato and HERA respectively. We also show that Rubato outperforms HERA when implemented in hardware in both performance and area.

\balance 
\bibliographystyle{IEEEtran}
\bibliography{main}

\end{document}